\documentclass[10pt,a4paper,twoside]{article}

\usepackage{amssymb,amsmath,multirow, graphics}

\newtheorem{theorem}{Theorem}[section]

\newtheorem{example}[theorem]{Example}

\newtheorem{lemma}[theorem]{Lemma}

\usepackage[dvips]{graphicx}
\usepackage{epsfig}

    \makeatletter
    \let\@fnsymbol\@arabic
    \makeatother

\title{Asymptotic Efficiency of New Exponentiality Tests Based on a Characterization  }

\author{Bojana Milo\v sevi\'c}

\begin{document}
\maketitle
\begin{abstract}
Two new tests for  exponentiality, of integral- and Kolmogorov-type, are proposed. They are based on a recent characterization and  formed using appropriate V-statistics. Their asymptotic properties are examined and their local Bahadur efficiencies  against some common alternatives are found. A class of locally optimal alternatives for each test is obtained.
The powers of these tests, for some small sample sizes, are compared with  different  exponentiality tests.
\end{abstract}
keywords:
\texttt{testing of exponentiality, order statistics, Bahadur\\ efficiency, $U$-statistics}


\section{Introduction}
Exponential distribution is probably one of the most applicable distribution in reliability theory, survival analysis
 and many other fields. Therefore, ensuring that  the data come from exponential  family of distributions is of a great importance.
 Goodness of fit testing for exponentiality has been popular for decades, and in recent times tests based on characterizations have become one of the primary  directions.
 Many interesting characterizations of exponential distribution can be found  in \cite{Ahsan}, \cite{arnold}, \cite{balakrishnan} and \cite{Galambos}. Goodness of fit tests based on characterizations of exponential distribution are studied in papers \cite{AA},
\cite{Ang}, \cite{Kou}, \cite{Kou2}, among others. In particular, the Bahadur efficiency of such tests has been considered in,  e.g.,  \cite{Niknik}, \cite{NikVol}, \cite{volkova}.


Recently Obradovi\'c \cite{obradovic} proved three new characterizations  of exponential distribution based on order statistics in small samples. In this paper we propose two new goodness of fit tests
based on one of those characterizations:

\textit{
Let $X_0,X_1,X_2,X_3$ be independent and identically distributed non-negative  random variables (i.i.d.) from the distribution whose density $f(x)$ has the Maclaurin's expansion for $x>0$. Let $X_{(2;3)}$ and $X_{(3;3)}$ be the median and maximum  of $\{X_1, X_2, X_3\}$.  If
\begin{equation}
 X_0 + X_{(2;3)}\overset{d}{=} X_{(3;3)}
\end{equation}
then $f(x)=\lambda e^{-\lambda x} $ for some $\lambda>0$.
}

\bigskip

Let $X_1,X_2,\ldots,X_n$ be i.i.d.
observations having the continuous d.f. $F$. We   test the composite hypothesis that $F$ belongs to family of exponential distributions $ {\mathcal{E}(\lambda)})$, where $\lambda>0$ is an unknown parameter.

We shall consider integral and Kolmogorov-type test statistics  which are invariant with respect to
the scale parameter $\lambda$ ( see \cite{Kor}):
 \begin{equation}
 I_n=\int\limits_{0}^{\infty}\int\limits_{0}^{\infty}(H_n(t)-G_n(t))dF_n(t),
\end{equation}
\begin{equation}
K_n=\sup\limits_{t\geq0}\big|H_n(t)-G_n(t)\big|,
\end{equation}
where $G_n$ and $H_n$ are $V$ empirical d.f.'s
\begin{align*}
 G_n(t)&=\frac{1}{n^4}\sum\limits_{i=1}^n\sum\limits_{j=1}^n\sum\limits_{k=1}^n\sum\limits_{l=1}^nI\{X_i+{\rm med}(X_j,X_k,X_l)<t\},\\
 H_n(t)&=\frac{1}{n^3}\sum\limits_{j=1}^n\sum\limits_{k=1}^n\sum\limits_{l=1}^nI\{\max(X_j,X_k,X_l)<t\}.
\end{align*}
In order to determine the quality of our tests and to compare them  with some other tests we shall use local Bahadur efficiency. We choose this type of asymptotic efficiency since it is applicable to non-normally distributed test statistics such as Kolmogorov. For asymptotically normally distributed test statistics local  Bahadur efficiency and classical Pitman efficiency coincide (see \cite{Wie}).

The paper is organized as follows. In section 2 we study the integral statistic $I_n$. We find its asymptotic distribution, large deviations and calculate its asymptotic efficiency
against some common alternatives. We also present a class of locally optimal alternatives. In section 3 we do the analogous study for Kolmogorov-type statistics.
In section 4 and 5 we do the comparison of our tests with some existing tests for exponentiality and give the real data example.

\section{Integral-type Statistic $I_n$}
The statistic $I_n$ is asymptotically equivalent to $U$-statistic with symmetric kernel (\cite{Kor})
\begin{eqnarray*}
&&\Psi(X_1,X_2,X_3,X_4,X_5)=\frac{1}{5!}\sum\limits_{\pi(1:5)}\big(I\{\max(X_{\pi_2},X_{\pi_3},X_{\pi_4})<X_{\pi_5}\}\\&-&
I\{X_{\pi_1}+{\rm med}(X_{\pi_2},X_{\pi_3},X_{\pi_4})<X_{\pi_5}\}\big),
\end{eqnarray*}
where $\pi(1:m)$ is the set of all permutations $\{\pi_1,\pi_2,...,\pi_m\}$ of set $\{1,2,...,m\}$.

Its  projection on $X_1$ under null hypothesis is
\begin{eqnarray*}
\psi(s)&=&E(\Psi(X_1,X_2,X_3,X_4,X_5)|X_1=s)\\&=&\frac{1}{5}\Big(P\{\max(X_2,X_3,X_4)<X_5\}-P\{s+{\rm med}(X_2,X_3,X_4)<X_5\}\Big)\\
&+&\frac{3}{5}\Big(P\{\max(s,X_3,X_4)<X_5\}-P\{X_2+{\rm med}(s,X_3,X_4)<X_5\}\Big)\\&+&
\frac{1}{5}\Big(P\{\max(X_3,X_4,X_5)<s\}-P\{X_2+{\rm med}(X_3,X_4,X_5)<s\}\Big).
\end{eqnarray*}
After some calculations we get
\begin{eqnarray}
\psi(s)=-\frac{1}{20}+\frac{2}{5}e^{-3s}-\frac{9}{10}e^{-2s}+\frac{1}{2}e^{-s}.
\end{eqnarray}
The expected value of this projection is equal to zero, while its variance is
\begin{equation*}
\sigma^2_{I}=E(\psi^2(X_1))=\frac{29}{42000}.
\end{equation*}
Hence this kernel is non-degenerate. Applying Hoeffding's theorem (\cite{Hoeffding}) we get that  the asymptotic distribution of $\sqrt{n}I_n$ is normal $\mathcal{N}(0,\frac{29}{1680})$.

\subsection{Local Bahadur efficiency}
One way of measuring the quality of the tests is calculating their  Bahadur asymptotic efficiency. This quantity can be expressed as the ratio of Bahadur exact slope, function  describing the rate of exponential decrease for the
attained level under the alternative, and double Kullback-Leibler distance between null and alternative distribution. More about theory on this topic can be found in   (\cite{Bahadur}, \cite{Nik}).

 According to Bahadur theory the exact slopes are defined  in the following way.
 Suppose that the sequence  $\{T_n\}$ of test statistics under alternative  converges in probability to some finite function $b(\theta)$.
 Suppose also that the following large deviations limit exists
 \begin{equation}\label{ldf}
  \lim_{n\to\infty} n^{-1} \ln
P_{H_0} \left( T_n \ge t  \right)  = - f(t)
 \end{equation}
  for any $t$ in an open interval $I,$ on which $f$ is
continuous and $\{b(\theta), \: \theta > 0\}\subset I$. Then the Bahadur exact slope is
\begin{equation}\label{slope}
c_T(\theta) = 2f(b(\theta)).
\end{equation}

The exact slopes always satisfy the inequality
\begin{equation}
\label{Ragav}
c_T(\theta) \leq 2 K(\theta),\, \theta > 0,
\end{equation}
where $K(\theta)$ is the Kullback-Leibler "distance" between the alternative $H_1$ and the null hypothesis $H_0.$

Given (\ref{Ragav}), the local Bahadur efficiency of the sequence of statistics ${T_n}$ is defined as
\begin{equation}\label{localBahadurEf}
e^B (T) = \lim_{\theta \to 0} \frac{c_T(\theta)}{2K(\theta)}.
\end{equation}

\bigskip

Let $G(\cdot,\theta)$, $\theta \geq 0$, be a family of d.f. with densities $g(\cdot,\theta)$, such that $G(\cdot,0)\in \mathcal{E}(\lambda)$ and the regularity conditions from (\cite{Nik}, Chapter 6), including differentiation along $\theta$ in all appearing integrals, hold. Denote  $h(x)=g'_{\theta}(x,0)$. It is obvious that
$\int_{0}^{\infty}h(x)dx=0$.

We now calculate the Bahadur exact slope for the test statistic $I_n$.
\begin{lemma}\label{largeDefiation f}
For statistic $I_n$ the function $f_{I}$ from \eqref{ldf} is analytic for sufficiently small $\epsilon>0$ and it holds
\begin{equation*}
f_{I}(\epsilon)=\frac{840}{29}\epsilon^2+o(\epsilon^2),\; \epsilon\rightarrow 0.
\end{equation*}
\end{lemma}
\textbf{Proof.} The kernel  $\Psi$ is bounded, centered and non-degenerate. Therefore we can apply the theorem of large deviations for non-degenerate $U$-statistics(\cite{nikiponi}) and get the statement of the lemma. \hfill{$\Box$}
\begin{lemma}\label{lemabinteg}For a given alternative density $g(x;\theta)$ whose distribution belongs
to $\mathcal{G}$
holds
\begin{equation}
b(\theta)= 5\theta\int\limits_{0}^{\infty}\psi(x)h(x)dx+o(\theta),\;\;\theta\to 0.
\end{equation}
\end{lemma}
\noindent\textbf{Proof}.
Using strong law of large numbers for $U-$ and $V-$ statistics we get that the $b(\theta)$ is
\begin{align}
\nonumber b(\theta)&=P\{\max(X_2,X_3,X_4)<X_5\}-P\{X_1+{\rm med}(X_2,X_3,X_4)<X_5\}\\
&=\frac{1}{4}-6\int\limits_{0}^{\infty}g(x,\theta)\int\limits_{0}^{\infty}G(y,\theta)(1-G(y,\theta))g(y,\theta)
\int\limits_{x+y}^{\infty}g(z,\theta)dz dy dx. \label{bInt}
\end{align}
Its first derivative along $\theta$ is
\begin{align*}
b_{\theta}'(\theta)&=-6\int\limits_{0}^{\infty}g'_{\theta}(x,\theta)
\int\limits_{0}^{\infty}G(y,\theta)(1-G(y,\theta))g(y,\theta)
\int\limits_{x+y}^{\infty}g(z,\theta)dz dy dx\\
&-6\int\limits_{0}^{\infty}g(x,\theta)\int\limits_{0}^{\infty}
G'_{\theta}(y,\theta)(1-2G(y,\theta))g(y,\theta)\int\limits_{x+y}^{\infty}g(z,\theta)dz dy dx
\\&-6\int\limits_{0}^{\infty}g(x,\theta)
\int\limits_{0}^{\infty}G(y,\theta)(1-G(y,\theta))g'_{\theta}(y,\theta)
\int\limits_{x+y}^{\infty}g(z,\theta)dz dy dx
\\&-6\int\limits_{0}^{\infty}g(x,\theta)
\int\limits_{0}^{\infty}G(y,\theta)(1-G(y,\theta))g(y,\theta)
\int\limits_{x+y}^{\infty}g'_{\theta}(z,\theta)dz dy dx.
\end{align*}
Letting $\theta=0$ we have
\begin{align*}
b_{\theta}'(\theta)&=-6\int\limits_{0}^{\infty}h(x)
\int\limits_{0}^{\infty}(1-e^{-y})e^{-2y}
\int\limits_{x+y}^{\infty}e^{-z}dz dy dx\\
&-6\int\limits_{0}^{\infty}e^{-x}\int\limits_{0}^{\infty}
H(y)(1-2(1-e^{-y}))e^{-y}\int\limits_{x+y}^{\infty}e^{-z}dz dy dx
\\&-6\int\limits_{0}^{\infty}e^{-x}
\int\limits_{0}^{\infty}(1-e^{-y})e^{-y}h(y)
\int\limits_{x+y}^{\infty}e^{-z}dz dy dx
\\&-6\int\limits_{0}^{\infty}e^{-x}
\int\limits_{0}^{\infty}(1-e^{-y})e^{-2y}
\int\limits_{x+y}^{\infty}h(z)dz dy dx,
\end{align*}
where $H(y)=G_{\theta}'(y,0)$.
 Transforming this expression we obtain
 \begin{equation*}
 b'_{\theta}(0)=\int\limits_{0}^{\infty}h(s)(-1 + 2 e^{-3 s} - \frac{9}{2} e^{-2 s} + \frac{5}{2} e^{-s})ds=5\int\limits_{0}^{\infty}h(s)\psi(s)ds.
 \end{equation*}

 Expanding $b(\theta)$ in Maclaurin series we get the statement of the lemma.
\hfill$\Box$

\bigskip

The  Kullback-Leibler "distance" from the alternative density $g(x,\theta)$ from $\mathcal{G}$ to the class of exponential densities $\{\lambda e^{-\lambda x},\;\;x\geq 0 \}$, is
\begin{equation}\label{kldef}
K(\theta) = \inf_{\lambda>0} \int_0^{\infty} \ln [g(x,\theta) / \lambda \exp(-\lambda x) ] g(x,\theta) \ dx.
\end{equation}
It can be shown (\cite{NikTchir}) that for small $\theta$ equation \eqref{kldef} can be expressed as
\begin{equation}\label{kul}
2K(\theta)=\bigg(\int\limits_{0}^{\infty }h^2(x)e^xdx-\Big(\int\limits_{0}^{\infty}xh(x)dx\Big)^2\bigg)\cdot \theta^2+o(\theta^2).
\end{equation}
This quantity can be easily calculated as $\theta \to 0$ for particular alternatives.

In what follows we shall calculate the local Bahadur efficiency of our test for some alternatives. The alternatives we are going to use are:
\begin{itemize}\label{alternative}
\item Weibull distribution with the density
\begin{equation}\label{vejbul}
g(x,\theta)=e^{-x^{1 + \theta}} (1 + \theta) x^{\theta},\;\theta > 0,\; x\geq 0;
\end{equation}
\item Makeham distribution with the density
\begin{equation}\label{makeham}
g(x,\theta)=(1+\theta(1-e^{-x}))\exp(-x-\theta( e^{-x}-1+x)),\;\theta > 0, \;x\geq 0;
\end{equation}
\item exponential mixture with negative weights (EMNW($\beta$)) \cite{vjevremovic} with density
\begin{equation}\label{mesavina} g(x,\theta)=(1+\theta)e^{-x}-\beta\theta e^{\beta x},\; \theta\in\big(0,\frac{1}{\beta-1}\big],\; x\geq 0;
\end{equation}
\item generalized exponential distribution (GED)(\cite{nadarajah}) with density
\begin{equation}\label{nagaraja}
g(x,\theta)=e^{1 - (1 + x)^{1 + \theta}} (1 + \theta) (1 + x)^{\theta}\;\theta > 0, x\geq 0;
\end{equation}
\item extended exponential distribution (EE) \cite{gomez}
\begin{equation}
\label{gomez}
g(x,\theta)=\frac{1 + \theta x}{1 + \theta}e^{-x},\; \theta > 0,\; x\geq 0.
\end{equation}
\end{itemize}

In the following two examples we shall present the calculations of local Bahadur efficiency.
\begin{example}
Let the alternative hypothesis be Weibull distribution with the density function \eqref{vejbul}.
The first derivative along $\theta$ of its density at $\theta = 0$ is
\begin{equation*}
h(x)=e^{-x} + e^{-x} \log x - e^{-x} x \log x.
\end{equation*}
\noindent Using \eqref{kul} we get that the Kullback-Leibler distance is $K(\theta)=\frac{\pi^2}{6}\theta^2+o(\theta^2),\;\;\theta\to 0.$
Applying lemma \ref{lemabinteg} we have
\begin{align*}
b_{I}(\theta)&=5\theta \int\limits_{0}^{\infty}\psi(x)(e^{-x} + e^{-x} \log x - e^{-x} x \log x)dx+o(\theta)\\&=\log\bigg(\frac{3}{2^{11/8}}\bigg)\theta+o(\theta)\approx 0.146\theta+
o(\theta),\;\;\theta\to 0.
\end{align*}
According to lemma \ref{largeDefiation f} and \eqref{localBahadurEf} we get that  local Bahadur efficiency $e_B(I)=0.746$.

\end{example}
The calculation procedure for alternatives (\ref{makeham}-\ref{nagaraja}) is analogous. Their efficiencies are given in table \ref{fig: LBEI}. The exception is the alternative \eqref{gomez}
where the lemma \ref{lemabinteg} and \eqref{kul} cannot be applied. We present it in the following example.
\begin{example} Consider the alternative (EE) with density function  \eqref{gomez}.
Its first derivative along $\theta$ at $\theta = 0$ is
\begin{equation*}
h(x)=-e^{-x} + e^{-x} x.
\end{equation*}
The expressions  $\int_{0}^{\infty} h(x)\psi(x)dx$ and
$\int_{0}^{\infty }h^2(x)e^xdx-\Big(\int_{0}^{\infty}xh(x)dx\Big)^2$ are equal to zero, hence we  need to expand the series for $b_{I}(\theta)$ and $2K(\theta)$ to the first non-zero term.
Limit in probability $b_I(\theta)$ from \eqref{bInt} is equal to
\begin{equation*}
b_I(\theta)=\frac{\theta^2 (168 + 356 \theta + 161 \theta^2)}{2304 (1 + \theta)^4}=\frac{7}{96}\theta^2+o(\theta^2),\;\;\theta\to 0.
\end{equation*}
The double Kullback-Leibler distance \eqref{kldef} from  \eqref{gomez} to family of exponential distributions is
\begin{equation}\label{KLgomezb}
2K(\theta)=2\frac{(1 + \theta) \log(
   1 + 2 \theta)-e^{1/\theta} \theta Ei(-(1/\theta)) }{(1 + \theta)^2}=\theta^4+o(\theta^4),\;\;\theta\to 0,
\end{equation}
where $Ei(z)=\int_{-z}^{\infty}\frac{1}{u}e^{-u}du$ is the exponential integral.
According to lemma \ref{largeDefiation f} and \eqref{localBahadurEf} we get that  local Bahadur efficiency $e_B(I)=0.481$.
\end{example}
\begin{table}[!hhh]\centering
\caption{Local Bahadur efficiency for the statistic $I_n$}
\bigskip
\begin{tabular}{|c|c|}
\hline
Alternative & Efficiency\\
\hline
Weibull & 0.746 \\
Makeham & 0.772 \\
EMNW(3) &0.916 \\
GED &0.556\\
EE&0.481\\\hline
\end{tabular}
\label{fig: LBEI}
\end{table}
We can notice from table \ref{fig: LBEI} that all efficiencies are reasonably high except in case of \eqref{gomez}, the example which was included to show the exception in calculation.

\subsection{Locally  optimal alternatives}
In this section we determine some of those alternatives for which statistic $I_n$ is locally asymptotically optimal  in Bahadur sense. More on
this topic can be found in \cite{Nik} and \cite{Nik1984}.
 We shall determine some of those alternatives in the
following theorem.
\begin{theorem}
Let $g(x;\theta)$ be the density from $\mathcal{G}$ that satisfies condition
\begin{equation}
\int\limits_{0}^{\infty} e^{x}h^2(x)dx<\infty.
\end{equation}
Alternative densities
\begin{equation*}
 g(x;\theta)=e^{-x}+e^{-x}\theta(C\psi(x)+D(x-1)),\; x\geq 0,\;C>0,\; D\in \mathbb{R},
\end{equation*}
are for small $\theta$ locally asymptotically optimal for the test based on $I_n$.
\end{theorem}
\noindent \textbf{Proof.}
Denote
\begin{equation}\label{h0}
 h_0(x)=h(x)-(x-1)e^{-x}\int\limits_{0}^{\infty} h(s)s ds.
\end{equation}
It is easy to show that this function satisfies the following equalities.

\begin{align}\label{H01}
 &\int\limits_{0}^{\infty} h_0^2(x) e^{x}dx=\int\limits_{0}^\infty e^{x}h^{2}(x)dx-
 \bigg(\int\limits_{0}^{\infty} h(x)x dx\bigg)^2\\
\label{HO2}
 &\int\limits_{0}^{\infty} \psi(x)h_0(x)dx=\int\limits_{0}^{\infty} \psi(x)h(x)dx.
 \end{align}
Local asymptotic efficiency is

\begin{align*}
 e^{B}_I&=\lim\limits_{\theta\rightarrow 0}\frac{c_{I}(\theta)}{2K(\theta)}=
 \lim\limits_{\theta\rightarrow 0}\frac{2f(b_{I}(\theta))}{2K(\theta)}=
 \lim\limits_{\theta\rightarrow 0}\frac{2\cdot\frac{840}{29}b^2_{I}(\theta)}{2K(\theta)}
=\lim\limits_{\theta\rightarrow 0}\frac{b_{I}^2(\theta)}{25\sigma_{I}^2 2K(\theta)}\\&=
\lim\limits_{\theta\rightarrow 0}
 \frac{25\theta^2\bigg(\int\limits_{0}^{\infty}\psi(x)h(x)dx\bigg)^2+o(\theta^2)}{25
 \int\limits_{0}^\infty \psi^2(x) e^{-x}dx \bigg(\theta^2(\int\limits_{0}^\infty e^xh^{2}(x)dx-
 \big(\int\limits_{0}^{\infty} h(x)x dx\big)^2)+o(\theta^{2})\bigg)}\\&=
 \frac{\bigg(\int\limits_{0}^{\infty}\psi(x)h(x)dx\bigg)^2}{
 \int\limits_{0}^\infty \psi^2(x) e^{-x}dx \bigg(\int\limits_{0}^\infty e^xh^{2}(x)dx-
 \big(\int\limits_{0}^{\infty} h(x)x dx\big)^2\bigg)}
 \\&=
\frac{\bigg(\int\limits_{0}^{\infty}\psi(x)h_0(x)dx\bigg)^2}{
 \int\limits_{0}^\infty \psi^2(x) e^{-x
 }dx \int\limits_{0}^{\infty} h^2_0(x) e^xdx }.
\end{align*}
From the Cauchy-Schwarz inequality we have that $ e^{B}_I=1$ if and only if
$h_0(x)=C\psi(x) e^{-x}$. Inserting that in \eqref{h0} we obtain $h(x)$.
The densities from the statement of the theorem have the same $h(x)$, hence the proof is completed.
$\hfill \Box$

\section{Kolmogorov-type Statistic $K_n$}

\begin{figure}[h!]
\begin{center}
\includegraphics[scale=0.8]{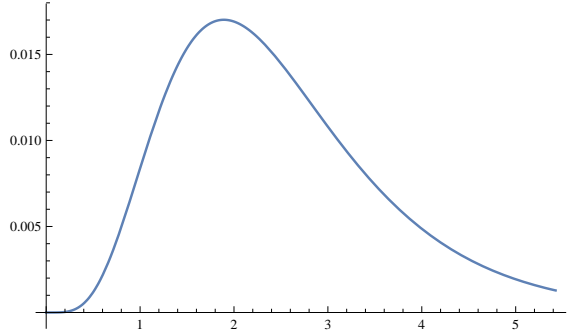}\caption{Plot of the function $\sigma_K^2(t),$  }
\label{fig: sigmaK1}
\end{center}
\end{figure}

For a fixed $t>0$  the expression $H_n(t)-G_n(t)$ is the V-statistic with the following kernel:
\begin{align*}
\Xi(X_1,X_2,X_3,X_4,t)&=\frac{1}{4!}\sum\limits_{\pi(1:4)}\Big(
I\{\max(X_{\pi_2},X_{\pi_3},X_{\pi_4})<t\}\\&-I\{X_{\pi_1}+{\rm med}(X_{\pi_2},X_{\pi_3},X_{\pi_4})<t\}\Big).
\end{align*}
The projection of this family of kernels on $X_1$ under $H_0$ is
\begin{align*}
\xi(s,t)&=E(\Xi(X_1,X_2,X_3,X_4,t)|X_1=s)
\\&=\frac{1}{4}\Big(P\{\max(X_2,X_3,X_4)<t\}-P\{s+{\rm med}(X_2,X_3,X_4)<t\}\Big)\\
&+\frac{3}{4}\Big(P\{\max(s,X_3,X_4)<t\}-P\{X_2+{\rm med}(s,X_3,X_4)<t\}\Big).
\end{align*}
After some calculations we get
\begin{align*}
\xi(s,t)&=
   \frac{1}{4}I\{s<t\}e^{-s-3t}(-e^s-2e^{4s}+6e^{2t}+3e^{s+t}+3e^{3s+t}-e^{s+2t}(9-6s))\\&+
   \frac{1}{4}I\{s\geq t\}e^{-3t}(-1+6e^t-2e^{3t}-e^{2t}(3-6t)).
\end{align*}
 The variances of these projections  $\sigma_K^2(t)$ under $H_{0}$ are
\begin{align*}
\sigma_K^2(t)&=\frac{9}{80}e^{-6t}-\frac{3}{8}e^{-5t}-\frac{3}{8}e^{-4t}-\frac{9}{8}e^{-3t}+\frac{33}{16}e^{-2t}-\frac{3}{10}e^{-t}.
\end{align*}
The plot of this function is shown in Figure \ref{fig: sigmaK1}.

We find that
\begin{equation*}
\sigma_K^2=\sup_{ t\geq0} \sigma_K^2(t)=0.017.
\end{equation*}
The supremum is reached for $t_0=1.892$.
Therefore,  our family of kernels $\Xi(X_1,X_2,X_3,X_4,t)$  is non-degenerate as defined in \cite{Niki10}.
It can be shown (see \cite{Silv}) that
$U$-empirical process
$$\sqrt{n} \left(H_n(t) - G_n(t)\right), \ t\geq 0,
$$
weakly converges in $D(0,\infty)$ as $n \to \infty$ to certain centered Gaussian
process  with calculable covariance.  Thus, the sequence of our test statistic $K_n$
 converges in distribution to the random variable   $\sup_{t\geq0} |\sqrt{n} \big(H_n(t) - G_n(t)\big)|$ but its distribution is unknown.

 Critical values for statistics $K_n$   for different sample size and the level of significance are shown in the table \ref{fig: critic}. They are calculated using Monte Carlo methods based on 10000 repetitions.

\begin{table}[!hhh]\centering
\bigskip
\caption{Critical values for statistics $K_n$ }
\bigskip
\begin{tabular}{|c|c|c|c|c|}
\hline
$n$ & $\alpha=0.1$ & $\alpha=0.05$ & $\alpha=0.025$ &$\alpha=0.01$\\
\hline
10&0.49& 0.56& 0.62& 0.70\\
20&0.33& 0.39& 0.43&0.48\\
30&0.26& 0.30& 0.34&0.38\\
40&0.23& 0.26& 0.29&0.31\\
50&0.20& 0.23& 0.25& 0.28\\
100 & 0.14& 0.16& 0.17&0.19 \\
\hline
\end{tabular}
\label{fig: critic}
\end{table}

\subsection{Local Bahadur efficiency}
The family of kernels  $\{\Xi_4 (X_1,X_2,X_3,X_4,t),\;t\geq 0\}$ is centered and bounded in the sense described in \cite{Niki10}. Applying the large deviation theorem for the supremum of the family of non-degenerate $U$- and $V$-statistics from \cite{Niki10} ,
we find function $f$ from \eqref{ldf}.
\begin{lemma}\label{large deviationK f}
For statistic $K_n$ the function $f_{K}$ from \eqref{ldf} is analytic for sufficiently small $\epsilon>0$ and it holds
\begin{equation*}
f_{K}(\epsilon)=\frac{1}{32\sigma^{2}_K}\epsilon^2+o(\epsilon^2)\approx1.84\epsilon^2+o(\epsilon^2) ,\; \epsilon\rightarrow 0.
\end{equation*}
\end{lemma}

\begin{lemma}\label{bkols} For a given alternative density $g(x;\theta)$ whose distribution belongs
to $\mathcal{G}$
holds\label{bK}
\begin{equation*}
 b_K(\theta) = 4\theta\sup\limits_{t\geq 0} \big| \int_{0}^{\infty} \xi (x;t)h(x)dx\big|+o(\theta), \, \theta \to 0.
 \end{equation*}
 \end{lemma}

\noindent Using Glivenko-Cantelli theorem for $V$-statistics \cite{helmersjanssen} we have
\begin{align}\label{bka}
b_K(\theta)&=\sup\limits_{t\geq 0} \big|P\{\max(X_2,X_3,X_4)<t\}-P\{X_1+{\rm med}(X_2,X_3,X_4)<t\}\big|
\\\nonumber&=\sup\limits_{t\geq 0} \big|G^3(t,\theta)-6\int\limits_{0}^tg(x,\theta)\int\limits_{0}^{t-x}G(y,\theta)(1-G(y,\theta))g(y,\theta)dydx\big|.
\end{align}
Denote
\begin{equation}\label{abk}
a(t,\theta)=G^3(t,\theta)-6\int\limits_{0}^tg(x,\theta)\int\limits_{0}^{t-x}G(y,\theta)(1-G(y,\theta))g(y,\theta)dydx.
\end{equation}
Performing calculations similar to those from lemma \ref{lemabinteg} we get
\begin{align*}
a_{\theta}'(t,0)&=\int\limits_{0}^{t}h(x)(4\xi(x,t)+e^{-3 t} (1 - 6 e^t + 2 e^{3 t} + e^{2 t} (3 - 6 t))dx
\\&=4\int\limits_{0}^{\infty}h(x)\xi(x,t)dx.
\end{align*}
Expanding $a(t,\theta)$ in Maclaurin series and inserting the result in \eqref{bka} we get the statement of the lemma.

Now we shall calculate the local Bahadur efficiencies in the same manner as we did for integral-type  statistic. For the alternatives \eqref{vejbul} and \eqref{gomez} the process of calculations is presented in following two examples, while for the others the values of efficiencies are presented in  table \ref{fig: LBEK}.
 \begin{example}
 Let the alternative hypothesis be Weibull distribution with density function \eqref{vejbul}.
 Using lemma \ref{bkols} we have
 \begin{equation}
 a(t,\theta)=4\theta\int_{0}^{\infty}\xi(x,t)(e^{-x} + e^{-x} \log x - e^{-x} x \log x)dx+o(\theta),\;\;\theta\to 0.
 \end{equation}
 The plot of the function function $a_{\theta}'(t,0)$,  is shown in figure \ref{fig: bKvejbul}.
 \begin{figure}[h!]
\begin{center}
\includegraphics[scale=0.8]{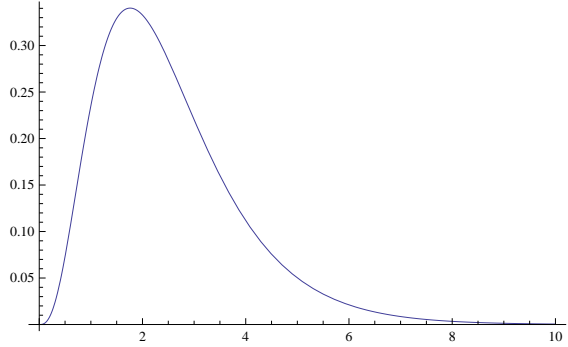}\caption{Plot of the function  $a_{\theta}'(t,0)$  }
\label{fig: bKvejbul}
\end{center}
\end{figure}
Supremum of $a(t,\theta)$ is reached at $t_1=1.761$, thus $b_K(\theta)=0.34\theta+o(\theta),\;\;\theta\to 0$.

\noindent Using lemma \ref{large deviationK f} and equations \eqref{slope} and \eqref{localBahadurEf} we get that the local Bahadur efficiency in case of statistic $K_n$ is 0.258.
 \end{example}

\begin{example}
Let the alternative density function be \eqref{gomez}.
The function $a(t,\theta)$ from \eqref{abk} is equal to
\begin{align*}
a(t,\theta)&=\frac{3}{4 (1 + \theta)^4}
  e^{-3 t}
   \theta^2 \big(-2 - 8 \theta - 9 \theta^2 - 2 t^3 \theta^2 -
    4 t^2 \theta (1 + 2 \theta) \\&+
    8 e^t (1 + \theta) (2 + t + 6 \theta + 4 t \theta + t^2 \theta) -
    t (2 + 10 \theta + 13 \theta^2) \\&+
    e^{2 t} (-14 - 56 \theta - 39 \theta^2 + t (6 + 18 \theta + 11 \theta^2))\big)\\
    &=\frac{3}{2} (-e^{-3t} + 8 e^{-2t} - 7 e^{- t} - te^{-3t} + 4 e^{-2t} t +
     3 e^{-t} t) \theta^2+o(\theta^2).
\end{align*}
The plot of the function $a_2(t)$, the coefficient next to $\theta^2$, in the expression above is given in figure \ref{fig: gomezB}. Thus we have
 \begin{equation*}
 \sup_{t\geq 0}|a(t,\theta)|=0.241\theta^2+o(\theta^2),\; \theta \to 0.
 \end{equation*}
 The value of double Kullback-Leibler distance is given in \eqref{KLgomezb}. Using lemma \ref{large deviationK f} and equations  \eqref{slope} and \eqref{localBahadurEf} we get that the local Bahadur efficiency  is 0.213.

 \begin{figure}[h!]
\begin{center}
\includegraphics[scale=0.8]{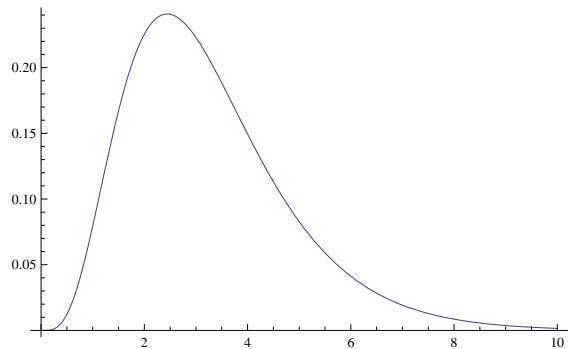}\caption{Plot of the function $a_2(t)$  }
\label{fig: gomezB}
\end{center}
\end{figure}
\end{example}
 \begin{table}[!hhh]\centering
\caption{Local Bahadur efficiency for the statistic $K_n$}
\bigskip
\begin{tabular}{|c|c|}
\hline
Alternative & Efficiency\\
\hline
Weibull &0.258  \\
Makeham & 0.370 \\
EMNW(3) &0.364 \\
GED&0.298\\
EE&0.213\\
\hline
\end{tabular}
\label{fig: LBEK}
\end{table}
We can see that, as expected, the efficiencies are lower than in case of the integral-type test. However the efficiencies are not that bad compared to some other Kolmogorov-type tests based on characterizations (e. g. \cite{volkova}).
 \subsection{Locally optimal alternatives}
 In this section we derive one class of alternatives that are locally optimal for test based on statistic $K_n$.
  \begin{theorem}
Let $g(x;\theta)$ be the density from $\mathcal{G}$ that satisfies condition
\begin{equation}
\int\limits_{0}^{\infty} e^{x}h^2(x)dx<\infty.
\end{equation}
Alternative densities
\begin{equation*}
 g(x;\theta)=e^{-x}+e^{-x}\theta(C\xi(x,t_0)+D(x-1)),\; x\geq 0,\;C>0,\; D\in \mathbb{R},
\end{equation*}
where $t_0=1.892$,
are for small $\theta$ locally asymptotically optimal for the test based on $K_n$.
\end{theorem}
\noindent \textbf{Proof.}
We use function $h_0$ defined in \eqref{h0}. It can be shown that function $h_0$ satisfies the condition \eqref{H01} and
\begin{equation*}
\int\limits_{0}^{\infty} \xi(x)h_0(x)dx=\int\limits_{0}^{\infty} \xi(x)h(x)dx.
\end{equation*}

Local asymptotic efficiency is

\begin{align*}
 e_{K}&=\lim\limits_{\theta\rightarrow 0}\frac{c_{K}(\theta)}{2K(\theta)}=
 \lim\limits_{\theta\rightarrow 0}\frac{2f(b_{K}(\theta))}{2K(\theta)}=
=\lim\limits_{\theta\rightarrow 0}\frac{b_{K}^2(\theta)}{16\sigma_{K}(t_0)^2 2K(\theta)}\\&=
\lim\limits_{\theta\rightarrow 0}
 \frac{16\theta^2\sup\limits_{t\geq 0}\bigg(\int\limits_{0}^{\infty}\xi(x,t)h(x)dx\bigg)^2+o(\theta^2)}{16
 \sup\limits_{t \geq 0}\int\limits_{0}^\infty \xi^2(x,t) e^{-x}dx \bigg(\theta^2(\int\limits_{0}^\infty e^xh^{2}(x)dx-
 \big(\int\limits_{0}^{\infty} h(x)x dx\big)^2)+o(\theta^{2})\bigg)}\\&=
 \frac{\sup\limits_{t\geq 0}\bigg(\int\limits_{0}^{\infty}\xi(x,t)h(x)dx\bigg)^2}{
 \sup\limits_{t\geq 0}\int\limits_{0}^\infty \xi^2(x,t) e^{-x}dx \bigg(\int\limits_{0}^\infty e^xh^{2}(x)dx-
 \big(\int\limits_{0}^{\infty} h(x)x dx\big)^2\bigg)}
 \\&=
\frac{\sup\limits_{t \geq 0}\bigg(\int\limits_{0}^{\infty}\xi(x,t)h_0(x)dx\bigg)^2}{
 \int\limits_{0}^\infty \xi^2(x,t) e^{-x
 }dx \int\limits_{0}^{\infty} h^2_0(x) e^xdx }.
\end{align*}
From the Cauchy-Schwarz inequality we have that $ e_{K}=1$ if and only if
$h_0(x)=C\xi(x,t_0) e^{-x}$. Inserting that in \eqref{h0} we obtain $h(x)$.
The densities from the statement of the theorem have the same $h(x)$, hence the proof is completed.
$\hfill \Box$

 \section{Power comparison}

For purpose of comparison we calculated the powers for  sample sizes $n=20$ and $n=50$ for some common distributions and compare results with some other tests for exponentiality which can be found in \cite{henze}. The powers are shown in tables \ref{fig: comparison20} and \ref{fig: comparison50}. The labels used are identical to the ones in \cite{henze}.
Bolded numbers represent cases where our test(s) have the higher or equal power than the competitors tests. It can be noticed that in majority of cases statistic $I_n$ is the most powerful. Statistic $K_n$ also in most cases performs better than the competitor tests for $n=20$, while it is reasonably competitive for $n=50$.
However there are few cases where the powers of both our tests  are unsatisfactory.

\section{Application to real data}
This data set represents inter-occurrence times of fatal accidents to British registered
passenger aircraft, 1946-63, measured in number of days and listed in the order
of their occurrence in time (see \cite{Pyke}):

\noindent 20 106 14 78 94 20 21 136 56 232 89 33 181 424 14\linebreak
 430 155 205 117 253 86 260 213 58 276 263 246 341 1105 50 136.\linebreak

\noindent Applying our tests to these data, we get the following values of test statistics $I_n$ and $K_n$, as well as the corresponding p-values:
\begin{center}
 \begin{tabular}{ccc}

 statistic &$I_n$ &$K_n $ \\\hline
  value &  0.04 &  0.21  \\
 p-value& 0.32 &  0.24 \\
\end{tabular}
\end{center}
so we conclude that the tests do not reject exponentiality.
\begin{table}[htbp]
\centering
\bigskip
\caption{Percentage of significant samples for different exponentiality tests $n=20$, $\alpha=0.05$ }
\bigskip
\centering
\begin{tabular}{ccccccccccc}
Alternative & $EP$ & $\overline{KS}$ & $\overline{CM}$ & $\omega^2$ & $KS$& $KL$ & $S$ &$CO$&$I$&$K$\\\hline
W(1.4)&36&35&35&34&28&29&35&37&\textbf{46}&32\\
$\Gamma(2)$&48&46&47&47&40&44&46&54&\textbf{59}&32\\
LN(0.8)&25&28&27&33&30&35&24&33&9&6\\
HN&21&24&22&21&18&16&21&19&\textbf{30}&\textbf{25}\\
U&66&72&70&66&52&61&70&50&\textbf{79}&\textbf{89}\\
CH(0.5)&63&47&61&61&56&77&63&80&23&20\\
CH(1.0)&15&18&16&14&13&11&15&13&\textbf{22}&\textbf{19}\\
CH(1.5)&84&79&83&79&67&76&84&81&22&20\\
LF(2.0)&28&32&30&28&24&23&29&25&\textbf{39}&\textbf{32}\\
LF(4.0)&42&44&43&41&34&34&42&37&\textbf{53}&\textbf{44}\\
EW(0.5)&15&18&16&14&13&11&15&13&\textbf{22}&\textbf{19}\\
EW(1.5)&45&48&47&43&35&37&46&37&\textbf{57}&\textbf{52}

\end{tabular}
\label{fig: comparison20}
\end{table}
\begin{table}[htbp]
\centering
\bigskip
\caption{Percentage of significant samples for different exponentiality tests $n=50$, $\alpha=0.05$ }
\bigskip
\centering
\begin{tabular}{ccccccccccc}
Alternative & $EP$ & $\overline{KS}$ & $\overline{CM}$ & $\omega^2$ & $KS$& $KL$ & $S$ &$CO$&$I$&$K$\\\hline
W(1.4)&80&71&77&75&64&72&79&82&\textbf{82}&62\\
$\Gamma(2)$&91&86&90&90&83&93&90&96&94&72\\
LN(0.8)&45&62&60&76&71&92&47&66&14&7\\
HN&54&50&53&48&39&37&54&45&\textbf{58}&50\\
U&98&99&99&98&93&97&99&91&\textbf{99}&\textbf{100}\\
CH(0.5)&94&90&94&95&92&99&94&99&41&37\\
CH(1.0)&38&36&37&32&26&23&38&30&\textbf{41}&\textbf{38}\\
CH(1.5)&100&100&100&100&98&100&100&100&40&38\\
LF(2.0)&69&65&69&64&53&54&69&60&\textbf{73}&62\\
LF(4.0)&87&82&87&83&72&75&87&80&\textbf{88}&79\\
EW(0.5)&38&36&37&32&26&23&38&30&\textbf{41}&37\\
EW(1.5)&90&88&90&86&75&79&90&78&\textbf{90}&88

\end{tabular}
\label{fig: comparison50}
\end{table}
\section{Conclusion}
In this paper  two goodness of fit tests based on a characterization were studied. The major advantage of our tests is that they are free of parameter $\lambda$.
The local Bahadur efficiencies  for some alternatives were calculated and the results are more than satisfactory.
For both tests  locally optimal class of alternatives were determined. 
These tests were compared with other   goodness-of-fit tests and it can be noticed that in most
cases  our tests are more powerful.

\end{document}